\documentclass[a4paper,12pt,onecolumn,final]{panl}
\usepackage{graphicx,amsmath,amssymb}
\usepackage{cite}
\def\be{\begin{eqnarray}}
\def\ee{\end{eqnarray}}

\newcommand{\lsim}{\stackrel{\scriptstyle <}{\phantom{}_{\sim}}}
\newcommand{\gsim}{\stackrel{\scriptstyle >}{\phantom{}_{\sim}}}

\originalTeX
\begin{document}\title{Charged pion vortices in rotating  systems}
\maketitle
\authors{ D.\,N.\,Voskresensky$^{\,a,b}$}
\from{$^{a}$BLTP, Joint Institute for Nuclear Research, RU-141980 Dubna, Russia}
\from{$^{b}$National Research Nuclear University ``MEPhI'', 115409 Moscow, Russia}



\begin{abstract}  Possibilities for  formation of the charged pion field vortices in a   rotating empty vessel (in vacuum) and  in the rotating pion   gas with a dynamically fixed particle number  at zero temperature are studied within the  $\lambda|\phi|^4$ model.  It is shown that in the former case at a rapid rotation a supervortex  of a charged pion field can be formed.  Important role played by  the electric field is demonstrated. Field configurations in presence and absence of the pion self-interaction are found.
 Conditions  for formation   of the vortex lattice at the rotation of the charged pion gas at zero temperature are studied.   Observational effects  are  discussed.
\end{abstract}
 \maketitle
\section{Introduction}
In heavy-ion collisions at some collision stage a hadron fireball    is formed. At  LHC and top RHIC collision energies  number of produced pions exceeds the baryon/antibaryon number  by an order of magnitude \cite{
Abelev,Adamczyk}.  At the fireball expansion  stage from chemical to thermal freeze out the pion number is approximately (dynamically) conserved. If the state formed at the chemical freeze-out was   overpopulated by pions, then during the cooling they may form the Bose-Einstein  condensate  characterized by the dynamically fixed pion number, as was suggested in   \cite{Voskresensky1994} and studied then in a number of works, cf. \cite{KKV1996,Voskresensky1996,Begun2015,Kolomeitsev:2019bju}.
The ALICE Collaboration observed a significant suppression of three and four pion
Bose–Einstein correlations in Pb-Pb collisions at
$\sqrt{s_{NN}} =2.76$ TeV at the LHC \cite{Abelev2014}.
Analysis \cite{Begun2015} indicated that about $5\%$ of pions could stem from the Bose-Einstein condensate.

Estimates show on angular
momenta $L\sim \sqrt{s}Ab/2\lsim 10^6\hbar$ in peripheral heavy-ion collisions of Au $+$ Au  at $\sqrt{s} = 200$ GeV, for the impact parameter $b = 10$ fm, where $A$ is the nucleon number of the ion  \cite{Xu-Guang Huang}. The global
polarization of $\Lambda (1116)$  hyperon observed  by the STAR Collaboration in non-central Au-Au collisions  \cite{Adamczyk} indicated existence of a vorticity with  rotation frequency  $\Omega\simeq
(9\pm 1) \cdot 10^{21}$ Hz $\simeq 0.05m_\pi$, $m_\pi\simeq 140$\,MeV is the pion mass.

Formation of vortices in  resting quantum liquids is energetically unfavorable. Vortex structures  in the rotating liquid helium and cold Bose gases   of nonrelativistic bosons have been extensively studied, cf. \cite{
Tilly-Tilly,
PitString}.

Besides a rotation, also strong magnetic fields are expected to occur at heavy-ion collisions and in compact stars. Estimates \cite{Voskresensky:1980nk} predicted values of the magnetic field up to  $\sim (10^{17}-10^{18})$G  for peripheral heavy-ion collisions at the energy $\sim$ GeV per nucleon. Also, fields  $H\lsim (10^{15}-10^{16})$G should exist at surfaces of magnetars, and m.b. still stronger fields in interiors.

Question about condensation of the noninterating charged pions in vacuum at a  simultaneous action of the rotation and a  strong magnetic field was  studied in  \cite{Zahed}.
Work \cite{Guo} included the pion self-interaction within the $\lambda |\phi|^4$ model and suggested  appearance of a giant pion vortex (supervortex).  Limit $H\to 0$  was not allowed and effects of the electric field  were disregarded. Reference \cite{Voskresensky:2023znr}   studied  possibilities of the appearance  of the  pion-$\sigma$ supervortex in the rapidly rotating empty vessel and in rotating nuclear systems, as well as  formation of the vortex lattice in the rotating pion gas at zero temperature, $T=0$, within the $\sigma$ model with taking into account the charge effects.
In the given paper these problems will be considered on example of the $\lambda |\phi|^4$  model. Units $\hbar=c=1$ will be used.

\section{Charged pion field in rotation  frame}\label{chir}
Let us  study behavior of the  charged pion vacuum  and a pion gas   at $T=0$  with a dynamically fixed particle
  number, in the rigidly rotating cylindrical system at the constant rotation frequency $\vec{\Omega}\parallel z$ in the cylindrical coordinates $(r,\theta, z)$, $\nabla =(\partial_r, \partial_\theta/r, \partial_z)$,  $r=\sqrt{x^2+y^2}$.   Interval in  rotation frame is  \cite{Chen2015},
$ (ds)^2=(1-\Omega^2 r^2)(dt)^2 +2\Omega y dx dt  -2\Omega x dy dt -(dr_3)^2\,,$
 $r_3= \sqrt{r^2 +z^2}$, with the tetrad   $e_0^t=e_1^x=e_2^y=e_3^z=1$, $e_0^x=y\Omega$, $e_0^y=-x\Omega$.  Other elements are zero;
 $e_\alpha =e_\alpha^\beta\partial_\beta$, and thereby $e_0=\partial_t +y\Omega\partial_x -x\Omega\partial_y$, $e_i =\partial_i$. Lattin index $i=1,2,3$, Greek index $\alpha,\beta =0,1,2,3$.
 Region $r>1/\Omega$ is beyond light cone.

In presence of the electromagnetic field $A^\alpha_{\rm lab}$  in the laboratory frame, the pion term in the Lagrangian density in the rotation frame renders \cite{Chen2015,Zahed,Guo}:
\begin{eqnarray}
&{\cal{L}}_{\pi}={|(D_t+y\Omega D_x -x\Omega D_y)\phi|^2} -{|D_i \phi|^2}-{m^{2}_\pi |\phi|^2}-\frac{\lambda |\phi|^4}{2}\,,\label{LVrot}
 \end{eqnarray}
where $D_\alpha =\partial_\alpha +ieA_\alpha$, $eA_\alpha=eA_\beta^{\rm lab} e^\beta_\alpha$, $e$ is the charge of the electron,  $\lambda$ is a positive constant.
Let us focus  on  the case $V_{\rm lab}=V_{\rm lab}(r)$ produced by the external charge density $n_p^{\rm lab}(\vec{r})$, $\vec{A}_{\rm lab}=0$
at $\Omega =0$.
 In the rotation frame we seek solution of the equation of motion in the form of the individual vortex with the center at $r=0$, cf. \cite{
 Tilly-Tilly}, $ \phi =\phi_{0}\chi (r)e^{i\xi (\theta)-i\mu t+ip_z z}\,,$
with $\phi_{0}=const$, $p_z =const$. Being interested in description of ground state we put $p_z=0$. Circulation of the $\xi(\theta)$-field,
$\oint d\vec{l}\,\nabla \xi =2\pi \nu\,,$
yields integer values of the winding number $\nu =0,\pm 1,...$ and $\nabla \xi =\nu /r$, $\xi =\nu\theta$,  thereby. The quantity $\mu$ has the sense of the energy of the ground state level of the $\pi^-$. In case of the pion gas at $T=0$ it coincides with the chemical potential.

Lagrangian density with  account of the rotation  can be presented as
\begin{eqnarray}
&{\cal{L}}_{\pi,V} ={\cal{L}}_{\pi}+{\cal{L}}_{V}\,,\quad
{\cal{L}}_{\pi}= {|\widetilde{\mu}\phi|^2 }-{|\partial_i\phi|^2}-{m^{\,2}_\pi |\phi|^2}-\frac{\lambda |\phi|^4}{2}\,,\label{LphiOmpiV}
\\
&{\cal{L}}_{V}=\frac{(\nabla V)^2}{8\pi e^2}+n_p V
\,,\quad \widetilde{\mu}=\mu  +\Omega\nu-V(r)\,. \nonumber\end{eqnarray}
The rotation term acts as a constant contribution to the electric potential for $r<1/\Omega$. Equation of motion  for the vortex field in the rotation frame is
\begin{eqnarray}
    [\widetilde{\mu}^2 +\Delta_r {{-\nu^2/r^2}} -m^{2}_\pi] \chi (r) -\lambda |\phi_{0}|^2\chi^3 (r)   =0\,,\label{piSigma1111pi}
  \end{eqnarray}
$\Delta_r=\partial_r^2 +{\partial_r}/{r}$, and the equation  for the electric field is
\begin{eqnarray}
\Delta V=4\pi e^2 (n_p-n_\pi)\,,\quad n_{\pi}=\frac{\partial{\cal{L}}_{\pi , V}}{\partial\mu}
=2\widetilde{\mu}|\phi|^2\,,\label{densmod1rot}
 \end{eqnarray}
 $n_{\pi}$ is the charged pion field (particle) density.
 Simplifying, we will  assume that $n_p$ is produced by very heavy particles (e.g., by protons being 7 times heavier than pions) and thereby we put $n_p\simeq n_p^{\rm lab}$ and  we take $n_p=n_p^0\theta (R-r)$, and either  $n_p^0=const>0$ or zero, and $\theta(x)$ is the step function, $R<1/\Omega$.

 The angular momentum associated with the charged pion field $\phi$  is
\begin{eqnarray}&\vec{L}_{\pi}=\int d^3 X [ \vec{r}_3\times\vec{P}_\pi]\,\,,\quad
{P}^i_\pi=T^{0i}_\pi=-\frac{\partial{\cal{L}}_{\pi}}{\partial\partial_t \phi}\nabla \phi -\frac{\partial{\cal{L}}_{\pi}}{\partial\partial_t{\phi}^*}\nabla \phi^*,\,\,\label{pphi}\end{eqnarray}
$T^{0i}_\pi$ is the $(0i)$ component of the energy-momentum. The energy density is
\begin{eqnarray}
&E_{\pi,V }=E_{\pi }+E_{V }=\mu n_\pi  -{\cal{L}}_{\pi , V}
\,.\label{EnerPhipi}\end{eqnarray}
Question arises what is distribution of vortices if they appear: a supervortex with the winding number $\nu\gg 1$ or the lattice of vortices with $\nu=1$ each?

\section{  Rotation and laboratory frames}\label{sect-rot-lab}

The question how to treat the rotating reference frame and the response  of the Bose field vacuum and the Bose gas at $T=0$ (i.e. the Bose-Einstein condensate) on the rotation in this frame is rather subtle due to necessity to fulfill the causality condition $r<1/\Omega$.   Thereby, we will associate the rotation frame with a rotating rigid body of a finite transversal size.
For instance, we may  consider either  vacuum or the pion gas  inside a  long empty cylindric vessel of a large internal transversal radius $R$, external radius $R_{>}$, hight $d_z\gg R$, a large  mass $M$  and constant mass-density $\rho_M$, rotating in the $z$ direction with constant cyclic frequency $\Omega$ at  $\Omega <\Omega_{\rm caus}=1/R_{>}$, as requirement of causality. Otherwise  solid vessel will be destroyed by  rotation.

In case of the vortex  placed in the center of the cylindric coordinate system $P_\theta = 2\widetilde{\mu} |\phi|^2\nu/r =n_{\pi} \nu /r$ and  using (\ref{pphi}) we have
$L^\pi_z =  2\pi d_z \int_0^R r dr  \nu n_{\pi}=\nu N_\pi$.

There are two possibilities: (1)  the charged pion system responses on the rotation creating the  vortex field in the rotation frame, and (2)  it does not rotate, cf. \cite{
Tilly-Tilly,
PitString}.
Further, there are  two possibilities: (i) conserving rotation frequency $\vec{\Omega}_{fin}=\vec{\Omega}_{in}$ of the rotating rigid body representing the rotation frame, and (ii) conserving angular momentum $\vec{L}_{fin}=\vec{L}_{in}$.

In case (i) the kinetic energy of the vessel    measured in the laboratory (resting) reference frame,
${\cal{E}}_{in}=\pi \rho_M \Omega_{in}^2 d_z (R_{>}^4 -R^4)/4$,
given for simplicity for  the nonrelativistic motion,
does not change with time, i.e. ${\cal{E}}_{in}=const$.
The loss of the energy due to a radiation  is recovered    from  an external source. We will consider situation when  in the laboratory frame the pion vortex field does not appear from  the vacuum. It is so if external fields (the electric field $V$ in our case) are not too strong. However there is still a possibility of the formation of the pion field from the vacuum in the rotation reference frame. In presence of  Bose excitations the final energy of the system is given by
$${\cal{E}}_{ fin}={\cal{E}}_{in}+{\cal{E}}_{\pi}[\Omega_{in}]\,,$$ ${\cal{E}}_{\pi}[\Omega_{in}]$ is the rotation part of the energy associated with a boson field in the  rotating system.  Note that in a deep electric potential well in absence of the material walls, instability for the creation  of $\pi^{\pm}$ pairs occurs, if the pion ground energy level, $\mu$, reaches $-m_\pi$.
In the rotating piece $r<R$ of the vacuum inside the vessel the pions  can be produced via reactions on the walls of the vessel already when the lowest energy level, $\mu$, reaches zero. Further, studying vacuum in empty rotating vessel we put $\mu=0$ whereas in a formal treatment of rotation frame one would put $\mu =-m_\pi$. The condition for the formation of the condensate in these cases is
 \be{\cal{E}}_{\pi}[\Omega_{in}]
 <0\,.\label{standEin-finRot}\ee
The same condition holds, if we deal with the pion gas with a dynamically fixed particle number, with the  difference that for the gas the chemical potential, $\mu >0$, is determined from condition of the  fixed particle number.

 In case (ii) the vessel is rotated owing to the  initially applied angular momentum $\vec{L}_{in}=\int d^3 X [ \vec{r}_3\times\vec{P}_{in}]$, $\vec{P}_{in}=\rho_M[\vec{\Omega}_{in},\vec{r}_3]$.
   The value $\vec{L}_{in}$ is conserved, provided one ignores a weak radiation, but it  can be redistributed between the massive vessel (stiff subsystem) and   the pion field (a softer subsystem),
     \begin{eqnarray}
  &\vec{L}_{in}=\vec{L}_{M,fin}+\vec{L}_{\pi}^{\rm lab}\,,\quad {\cal{E}}_{fin}=\pi \rho_M \Omega_{fin}^2 d_z (R_{>}^4 -R^4)/4 +{\cal{E}}_{\pi}^{\rm lab}\,,\label{Leq}\end{eqnarray}
and for the gas with fixed particle number we have
  ${\cal{E}}_{\pi}^{\rm lab}={\cal{E}}_{\pi}[\nu, \Omega =0]\,.$
Employing (\ref{Leq})
and
neglecting $O(1/M)$ term we obtain
\be \delta {\cal{E}}={\cal{E}}_{fin}- {\cal{E}}_{in}\simeq -L_{\pi}^{\rm lab}\Omega_{in}+ {\cal{E}}_{\pi}^{\rm lab}[\nu, \Omega =0]\,.\label{standEin-fin}\ee
The vortex-condensate field appears provided $\delta {\cal{E}}<0$.  For $V_0<m$, ${\cal{E}}_{\pi}[\nu, \Omega =0]>0$.   Conditions  (\ref{standEin-finRot}) and (\ref{standEin-fin}) should coincide, see below.


\section{Charged pion vortex field in absence of self-interaction}\label{sect-ideal}

{\em Equation of motion, boundary conditions,   energy in rotation frame.}
Let us consider the case $\lambda =0$.
From Eq. (\ref{piSigma1111pi})
we  have
\be
\left[\partial_r^2 +{\partial_r}/{r} -{\nu^2}/{r^2}+(\widetilde{\mu}^2-{m}^2)
\right]\chi(r)=0\,. \label{KGF}
\ee
Eq. (\ref{KGF}) describes a spinless relativistic particle of  energy $\epsilon_{n,\nu}=\mu$,  mass ${m}$ and $z$-projection of the angular momentum $\nu$, placed  in the    potential well $U(r)=- \Omega\nu +V(r)$  for $r<R$.  Behavior at $r>R$ depends on the  boundary condition put at $r=R$. We are interested in the description of the ground state, then ${\mu}=\mbox{min}\{\epsilon_{n,\nu}\}$ plays a role  of the chemical potential. The term   $-2\Omega\nu({\mu}-V) |\phi|^2$
 in the energy density is  associated with the Coriolis force and the term $-\Omega^2\nu^2 |\phi|^2$ is  an attractive relativistic  $\propto 1/c^2$ contribution to the centrifugal force term  $(\nu^2/r^2)|\phi|^2$.

  The
 Schr\"odinger equation for a nonrelativistic spinless particle follows from the Klein-Gordon equation (\ref{KGF}) after  replacement ${\mu}\to {m}+\mu_{\rm n.r.}$ and subsequent dropping of  small quadratic terms $(\mu_{\rm n.r.}+\Omega\nu -V)^2$. Then we get
 \be \left[-{\Delta_r}/{(2{m})} -\Omega\nu+V(r) +{\nu^2}/{(2{m}r^2)}\right]\chi =E_{\rm n.r.}\chi,\label{Schrod}\ee
  $U_{ef}=-\Omega\nu +V(r) +{\nu^2}/(2{m}r^2)$, $E_{\rm n.r.}=\mu_{\rm n.r.}$. So, rotation in the  rotating frame acts similarly  to a constant  electric potential acting on a nonrelativistic particle with the projection of the angular momentum $\nu$.

  In the Schr\"odinger equation the rotation is ordinary introduced employing the local Galilei transformation, with the speed given by $\vec{W}=[\vec{\Omega}\times \vec{r}_3]$. It results in the replacement in the Schr\"odinger equation, cf. \cite{FischerBaym2003},
\be -{\Delta }/{(2{m})}\to -{(\nabla -i {m}\vec{W})^2}/{(2{m})}-{{m}\vec{W}^2}/{2}\to -{\Delta }/{(2{m})}-\Omega\nu\,,
\label{rotSchrod}\ee
that yields the same Eq. (\ref{Schrod}), at which we arrived considering the problem in the rotation frame. Thus we see that uniform rotation acts in nonrelativistic case  similarly to  a uniform rather weak magnetic field described by the vector-potential $\vec{A}=\frac{1}{2}[\vec{H},  \vec{r}_3]$. In  relativistic case shift of variables $\partial_t\to \partial_t-\Omega \partial_\theta$
 in the Klein-Godron equation in the rotation frame is not equivalent to the shift  $\nabla\to \nabla -i{m}[\vec{\Omega},\vec{r}_3]$ in the Hamiltonian and  subtraction of the $\frac{{m}\vec{W}^2}{2}$ term associated with the motion of the system as a whole, cf.  \cite{Chen2015}.

Further let us for simplicity consider the case $V\simeq -V_0=const$ for $r<R$. $V_0$  can be treated as a contribution to the chemical potential.
For example we may assume that  an ideal rotating vessel is placed inside the cylindrical co-axial charged capacitor or itself it  represents the capacitor.  Appearance of the field $\phi\neq 0$ produces a dependence of $V$ on $r$. However, if  $\phi$ is a rather small, we can continue to consider $V= -V_0\simeq const$. Employing dimensionless variable $x=r/r_0$, with
\be r_0=1/\sqrt{\bar{\mu}^2-{m}^2}\,, \,\quad \bar{\mu}=\mu+\Omega\nu+V_0\,,
\label{rlin}\ee
for $\bar{\mu}> {m}$,  $V_0=const$, from Eq. (\ref{KGF}) we obtain equation
 \be (\partial^2_x +x^{-1}\partial_x -\nu^2/x^2)\chi+\chi =0\,.\label{filvorteqdim1}
 \ee

Simplest appropriate boundary conditions are
\be\chi (0)=0\,,\quad \chi (R/r_0)=0\,.\label{boundcon}\ee

Further to be specific let us consider $\Omega, \nu>0$.
 Appropriate solution of Eq. (\ref{filvorteqdim1})  is the Bessel function
$\chi (r)=J_\nu (r/r_0)$
  for $\nu >0$,
  cf. \cite{Gradshtein}. For $x\to 0$ we have $J_\nu \sim x^\nu$. The energy of the $n=1$ level is determined by the first zero of the function  $J_\nu (R/r_0)=0$, $j_{n=1,\nu =0}\simeq 2.403$. The  $n=1,\nu=1$ zero yields  $j_{1,1}=R/r_0\simeq 3.832$,  $j_{1,\nu}$ increases with increase of integer values of $\nu$.
For $\nu\gg 1$, $j_{1,\nu}^{\rm as}\to \nu+1.85575\nu^{1/3}$, e.g., $j_{1, 100}\simeq 108.84$.

 Employing the boundary condition $\chi(x=R/r_0=j_{n,\nu})=0$
 we find
 \begin{eqnarray} &\epsilon_{n,\nu}=\mu=-\Omega\nu -V_0
 +{m}\sqrt{1+j_{n,\nu}^2/(R^2 {m}^2)}\,,
 \label{grlev}\end{eqnarray}
 $\bar{\mu}=\sqrt{{m}^2+j_{n,\nu}^2/R^2}.$
Notice that with increasing quantity $\Omega\nu +V_0$
the $n,\nu\neq 0$ levels become more bound than the level $n=1,\nu=0$.
From (\ref{grlev}) we  also find that the roots  $\epsilon_{n,\nu}$ for $V_0=0$ do not reach zero for
 $\Omega R<1$. Thus   for $V_0,\lambda=0$ the field $\phi$ would not appear at the rotation of empty vessel.

One could  employ the boundary condition $\chi^{\prime}(r=R)=0$ instead of the  condition $\chi(r=R)=0$. In both cases there is no current through the surface $r=R$. Such a change of the boundary condition would not affect our conclusion that at $V_0=0$ the energy level does not cross zero. Note that in cases of the vacuum and the Bose-Einstein condensate in the vessel, usage of one of mentioned boundary  conditions is motivated provided the typical frequency of atomic transitions in the solid wall, $\omega_{\rm at}$, is larger than difference between energies of  the first excited energy levels and the ground state level, $\sim 1/R$ for $\nu\gg mR$ and $\sim \nu/(R^2m)$ in oppisite case.  Otherwise one should use exact matching conditions for $\chi (r=R)$ and $\chi' (r=R)$.

{\em Storm of charged pion field in rotation frame.}
 The value of the  rotation frequency, at which Eq. (\ref{grlev}) could be fulfilled for $\mu =\epsilon_{1,\nu}\leq 0$, $V_0\neq 0$ is
\begin{eqnarray} &\Omega\geq \Omega_c=\Omega (\epsilon_{1,\nu}=0)=(\sqrt{m^2+j_{1,\nu}^2/R^2}-V_0)/\nu\,,\label{critOmvac}\end{eqnarray}
where $\sqrt{m^2+j_{1,\nu}^2/R^2}\to \sqrt{m^2+{\nu}^2/R^2}\,\, {\rm for}\,\,\nu\gg 1\,.$

For $1\leq \nu =c_1 {m}R\ll {m}R$, i.e. at $c_1\ll 1$,  ${m}R\gg 1$, from Eq. (\ref{grlev}) we get
\be
\epsilon_{1,\nu}\simeq -V_0-\Omega\nu +{m}+...\label{largenulimless}\ee
The levels $\epsilon_{1,\nu}$ reach zero  for  $V_0> {m}(1-c_1)$ at $\Omega =1/R$. The critical rotation frequency is then given by
\be\Omega_c =\Omega (\epsilon_{1,\nu}=0, c_1\ll 1)\simeq ({m}-V_0)/\nu >0.\label{Omcrsmall}
\ee
For $V_0<m$, $\Omega_c$ decreases with increasing $\nu$. For $V_0>m$, $\Omega_c\sim O(1/(mR^2))$. The latter case is similar to the case of the gas with fixed particle number.

For $\nu=c_1{m} R\gg {m} R\gg 1$ (at $c_1\gg 1$) from (\ref{grlev}) we have
\begin{eqnarray}
&\epsilon_{1,\nu}\simeq -V_0+(-\Omega R +1)\nu/R+R{m}^2/(2\nu)+1.86\nu^{1/3}/R+...\label{largenulim}
\end{eqnarray}
Setting in (\ref{largenulim})  the  limiting value $\Omega^{\rm caus}=1/R$ we see that $\epsilon_{1,\nu}\to -V_0+{m}/(2c_1)+...$  for $1\ll c_1\ll \sqrt{mR}$. Thus the level $\epsilon_{1,\nu}$ may reach zero for $\Omega R<1$ at $V_0>V_{0c}={m}/(2c_1)$. With increasing 
$\nu$,
$V_{0c}$
decreases.
Formation of the supervortex  becomes energetically favorable  for $\sqrt{mR}\gg c_1\gg 1$ at
\begin{eqnarray}
&\Omega>\Omega_c =\Omega (\epsilon_{1,\nu}=0,c_1\gg 1)\simeq \frac{1}{R}-\frac{V_0-{m}/(2c_1)}{c_1{m}R}\,,\quad V_0>V_{0c}=\frac{m}{2c_1}\,. \label{condVom}\end{eqnarray}
 In case (i) the minimal critical value $V_{0c}\sim \sqrt{m/R}$ for $c_1\sim \sqrt{mR}\gg 1$. For $V_0=V_{0c}$, we have $\Omega_c\to 1/R$. For $c_1\gg \sqrt{mR}$, $V_{0c}\simeq 1.86 (m/R^2)^{1/3}\ll m$. The amplitude of the arising vortex field is limited by  redistribution of the charge, which we did not take into account assuming that $V_0\simeq const$. In case (ii) $c_1$ is also limited by conservation of the initial angular momentum.

In the rotation frame, in case (i),  using solution (\ref{grlev}) and (\ref{EnerPhipi}), (\ref{filvorteqdim1})
we find
\begin{eqnarray}
&{\cal{E}}_{\pi} (\Omega)=\epsilon_{n,\nu}N_\pi =\epsilon_{n,\nu}d_z 4\pi\int_0^R r dr  \sqrt{m^2_\pi+\frac{j_{n,\nu}^2}{R^2} }\phi_{0}^2\chi^2(r)
.
\label{bosenrot}
\end{eqnarray}
For $n_p\neq 0$,  ${\cal{E}}_{\pi,V}\simeq {\cal{E}}_\pi (\Omega) +V_0 Z\,,$ where $Z=n_p \pi R^2 d_z$.

In case (ii) the angular momentum needed for formation of the vortex is taken from the bucket walls.
Presenting  Eq. (\ref{grlev}) as $\epsilon_{n,\nu}=\epsilon_{n,\nu}[\Omega =0]-\Omega\nu$ and comparing   Eqs. (\ref{standEin-fin}) and (\ref{bosenrot}) we get $\delta{\cal{E}}={\cal{E}}_{\pi} (\Omega)=\epsilon_{n,\nu}N_\pi$.

Notice that in case of the vacuum placed in a strong static electric field in absence of the rotation, the charged bosons
are  produced nonlocally  via tunneling  from the lower continuum to the upper continuum. The typical time of such processes is exponentially large $\tau\sim e^{m^2/|eE|}/ m$ for the strength of the electric field $|eE|\ll m$. In case of the rotating empty vessel the charged pion field can be produced in more rapid processes, in reactions of particles of the rotating wall of the vessel.
Also, as one of possibilities to create  the vortex field, one may inject  inside the  vessel an admixture of protons. Accelerated protons will then produce radiation of the charged pion pairs and the latter can then form the vortex field.

{\em Ideal pion gas with dynamically fixed particle number in rotating system.}
In case of the  ideal pion gas at $T=0$ characterized by the dynamically fixed particle number  $N_\pi$, being put in a resting vessel on the ground state level,
the value $\phi^2_{0}[\Omega =0]$  is found from the normalization condition $N_\pi={\mu} \phi^2_{0}[\Omega =0]\pi R^2d_z$. In the rotation frame $\epsilon_{1,\nu}=\mu >0$,
the constant $\phi^2_{0}[\Omega]$  is found from the  condition (\ref{bosenrot}) yielding
 $N_\pi
 \simeq 2\bar{\mu}\phi_{0}^2[\Omega] \pi  R^2 d_z J_{\nu+1}^2 (R/r_0[\bar{\mu}])\,,$
 with $r_0$ and $\bar{\mu}$  given in  Eq. (\ref{rlin}), i.e. the relation between the (dynamically) fixed  value $N_\pi$ and constant $\bar{\mu}$. Value  $\epsilon_{n,\nu}={\mu}$ depends on $\Omega$ through the relation (\ref{rlin}).

  To understand will the gas be at rest or rotating with the angular velocity $\Omega$ we should compare ${\cal{E}}_{\pi}[\Omega]=N_\pi \epsilon_{1,\nu}$ and ${\cal{E}}_{\pi}[\Omega=0]$. The minimal value of  ${\cal{E}}_{\pi}[\Omega=0]$ corresponds to $\nu =0$ and for ${m}_\pi R\gg 1$ is given by
\begin{eqnarray}{\cal{E}}_{\pi}[\Omega=0]\simeq N_\pi[-V_0+{m}_\pi+j^2_{1,0}/(2R^2{m}_\pi)]\,.\end{eqnarray}
For $\nu\ll R{m}_\pi$ and any $V_0$
we find that for
 \be\Omega>\Omega_{c1}^{\rm id}(\nu)\simeq  (j^2_{1,\nu}-j^2_{1,0})/(2\nu R^2 {m}_\pi)
 \,,\label{Omcr1id}
\ee
\begin{eqnarray}
&{\cal{E}}_{\pi}[\Omega]-{\cal{E}}_{\pi}[\Omega=0]
\simeq N_\pi [-\Omega\nu +(j^2_{1,\nu}-j^2_{1,0})/(2R^2{m}_\pi)]<0\,,
\end{eqnarray}
 $\Omega_{c1}^{\rm id}(\nu)$ is minimal for  $|\nu|=1$.  Comparing (\ref{Omcr1id}) and (\ref{critOmvac}), (\ref{Omcrsmall})
we see that $\Omega_{c1}^{\rm id}\ll \Omega_{c}(\lambda =0)$ at least for small values  $V_0$, i.e. in presence of a pion gas vortices appear already at much smaller rotation frequencies than in case of the rotating vacuum. For the former case at ${m}_\pi R\gg 1$ we deal with a slow rotation for $\Omega\sim \Omega_{c1}^{\rm id}(1)$. For a single vortex with $\nu=1$
we have
$\delta {\cal{E}}^{(1)} \simeq -  [\Omega-\Omega_{c1}^{\rm id}(1)]N_\pi\,.$
For $\nu_{\rm tot}$ single vortices, each  with $\nu=1$, the energy gain is $\delta {\cal{E}}=\nu_{\rm tot} \delta {\cal{E}}^{(1)}$.
For $\Omega > \Omega_{c1}^{\rm id}$ at  increasing rotation frequency   individual vortices may form lattice. For $\Omega<\Omega_{c1}$,  production of vortices is energetically not profitable. However, if a vortex appeared  by a reason, it would survive   due to conservation of $\nu$.

\section{Self-interacting complex scalar field in  rotating system}\label{sect-selfinter}

{\em Equation of motion, boundary conditions, energy.}
For $\lambda\neq 0$, $V_0\simeq const$, using Eq. (\ref{piSigma1111pi})
in the dimensionless variable $x=r/r_0$,
 we arrive at equation:
\be (\partial^2_x +x^{-1}\partial_x -\nu^2/x^2)\chi+\chi -\lambda\phi_{0}^2r_0^2\chi^3=0\,,\label{filvorteqdim11lambda}
\ee
where $r_0$ is given by Eq. (\ref{filvorteqdim1}).  Choosing $\lambda\phi_{0}^2r_0^2=1$ we have
\begin{eqnarray}&\phi_{0} =\sqrt{ (\bar{\mu}^2-{m}^{2}_\pi)/\lambda}\times\theta ((\bar{\mu}^2-{m}^2_\pi)/\lambda),\label{Phich2}
\end{eqnarray}
being the solution of Eq. (\ref{filvorteqdim11lambda}) at $x\to \infty$ satisfying condition $\chi (x\to \infty)\to 1$.

 For case of the rotating vacuum, for $\Omega -\Omega_c^\pi\sim \Omega_c^\pi$, $V_0\ll m_\pi$, cf. Eq. (\ref{omcr1mod1}) below,  we have $r_0\ll R$ for $m_\pi R\gg 1$.
 For the pion gas at a low density and small  $\Omega$ and $V_0$, from Eq. (\ref{densmod1rot}) it follows that $n_\pi\simeq 2\bar{\mu}(\bar{\mu}^2-m^2_\pi)/\lambda$, and we have  $\bar{\mu}\simeq m_\pi +n_\pi\lambda/(4m^2_\pi)$  for $n_\pi\ll m^3_\pi/\lambda$ and
thus $r_0 \simeq  \sqrt{2m_\pi/(n_\pi\lambda})$ in this case and $r_0\ll R$ for $m_\pi R\gg 1$ and $r_0\ll R/\nu$ for $c_1\ll 1$.

 There exist  asymptotic solution of Eq. (\ref{filvorteqdim11lambda}):  $\chi \propto x^{|\nu|}$ for $x\to 0$ and  $\chi =1-\nu^2/(2 x^2)$ for $x\gg \nu$.  Then  the field $\phi$ is expelled from the vortex core and the equilibrium value (\ref{Phich2})
 is recovered at $r\gg \nu r_0$. The solution is modified for $R-r\sim r_0$ to fulfill condition $\chi (R/r_0)=0$.

 To clarify how to fulfill the boundary condition at $r=R$     we can solve Eq. (\ref{filvorteqdim11lambda})   employing  the variable $y=(r-R)/r_0$,  $x= y+R/r_0$,  now at  boundary conditions
$\chi (y\to -\infty)=1$ and $\chi (y=0)=0$.
At $\nu r_0\ll  R$ for typical dimensionless distances $y\sim 1$, the angular momentum  term, $\sim \nu^2 r_0^2 /R^2\ll 1$, and  the curvature term, $\sim r_0/R\ll 1$, can be dropped, which  means that  geometry can be considered as effectively one-dimensional one, cf. a similar argumentation employed in \cite{Migdal:1977rn}.
Then appropriate solution gets  the form
\be \phi =-\phi_{0} e^{i\nu\theta}\mbox{th}[(r-R)/(\sqrt{2} r_0)], \,\, \nu r_0\ll R\,, r<R\,.\label{tangPhi}
\ee

For $V_0\simeq const$  and for $ (\bar{\mu}^2-{m}^2_\pi)/\lambda>0$ using (\ref{EnerPhipi}), (\ref{LphiOmpiV}), (\ref{filvorteqdim11lambda}) we find
\begin{eqnarray}
&{\cal{E}}_{\pi,V}\simeq
2\pi d_z \int_0^R r dr\left[-\frac{(\bar{\mu}^2-{m}_\pi^2)^2\chi^4}{2\lambda}
+\frac{2\mu\widetilde{\mu}
(\bar{\mu}^2-{m}^2_\pi)\chi^2}{\lambda}+n_p V_0\right].\label{Egcor1}\end{eqnarray}

{\em Rotating vacuum.}
In this case $\mu =\bar{\mu} -\Omega\nu - V_0=0$.
For $n_p=0$  only first term in square brackets (\ref{Egcor1})
remains. So,  production of the vortex field becomes   energetically favorable, ${\cal{E}}_{\pi,V}(\Omega)<0$, for $\Omega\nu+V_0>{m}_\pi$, i.e. for
\be\Omega>\Omega_c^{\pi} =({m}_\pi-V_0)/\nu\,, \quad c_1>({m}_\pi-V_0)/m_\pi >0\,.\label{omcr1mod1}\ee
Note that  $\Omega_c^{\pi}$ approximately coincides with $\Omega_c$ given above by Eq. (\ref{Omcrsmall}) derived for $\lambda =0$ and $c_1\ll 1$, but differs from (\ref{condVom}). On the other hand  the asymptotic solution $\chi =1-\nu^2/(2x^2)$  works only for $R\gg \nu r_0$, being valid for $\Omega -\Omega_c^{\pi}\sim \Omega_c^{\pi}$, $c_1\ll 1$, as well as  for $\Omega$ near $1/R$ and $V_0\gg V_c$.

{\em Non-ideal gas with fixed particle number in rotating system.}
Now let us consider non-ideal pion gas at $T=0$ with dynamically fixed particle number at the condition $\mu \simeq m+O(n_\pi/m^2)\gg \Omega\nu+V_0$. This case is similar to that occurs for cold atomic gases, and He-II when $\mu\simeq m_{\rm He}$ and $\Omega\nu+V_0\ll m_{\rm He}$.


In presence of the rotation, in the rotation frame, using  the asymptotic solution $\chi =1-\nu^2/(2 x^2)$ of the equation of motion for $x\gg \nu$ we find that the energy balance  is controlled by the kinetic energy of the vortex, ${\cal{E}}_{\rm kin}^{(1)}$, and the rotation contribution $L\Omega$ extracted from the first two  terms in squared brackets (\ref{Egcor1}). The same consideration can be performed in the laboratory frame employing Eq. (\ref{standEin-fin}). In the latter case  the  kinetic energy associated with the single vortex line  with the logarithmic accuracy is given by
\be{\cal{E}}_{\rm kin}^{(1)}\simeq  {\int d^3 X |\nabla \phi|^2} =\ln (\widetilde{R}/{r_0}){d_z \pi \nu^2 n_{\pi}}/{\bar{\mu}}
.\label{Evortex1}\ee
At large distances $r$ we cut   integration at $r\sim \widetilde{R}\gg r_0$, $\widetilde{R}$ being   the transversal size of the vessel $R$ in case of the single vortex line with the   center at $r=0$, and  the distance $R_L$ between vortices in case of the lattice of vortices. 
At small distances  integration is naturally cut at $r\sim r_0$.

Let us consider  the system at approximately constant density $n_{\pi}$.  Then from the condition ${\cal{E}}_{\rm kin}^{(1)}-\vec{L}\vec{\Omega} <0$  we  find that the first vortex filament appears (together with the anti-vortex)  for
\be
\Omega >\Omega_{c1}^\lambda (\nu)={\nu\ln ({R}/{r_0})}/{(R^2 \bar{\mu})}
\,,\label{Omcr1}
\ee
and for a low density and  slowly rotating gas $\bar{\mu}\simeq \mu\simeq {m}+O(n_\pi /m^2)$.
Note that Eq. (\ref{Omcr1})  differs only by a logarithmic pre-factor from a similar expression Eq. (\ref{Omcr1id}) valid for  the ideal gas.
We should put $\nu =1$ and take $\widetilde{R}\sim R$,
  getting the minimum value of $\Omega_{c1}^\lambda$. In case $R\gg r_0$,  following (\ref{Omcr1}) we have  $\Omega_{c1}^\lambda R\ll 1$ for a slow rotation, $c_1\ll 1$.


Under action of fluctuations the vortex lines may form spirals and rings, cf. \cite{PitString,Voskresensky:2023znr}.
Vortices may also form a   lattice and then the system
mimics   rotation of  the rigid body characterized by the linear velocity $v_{\rm rig}=\Omega R<1$.
In  case of a vortex lattice we get \cite{Tilly-Tilly,Voskresensky:2023znr}:
\be N_v^{\rm rig} \kappa = n_v \pi R^2 \kappa =2\pi R\cdot \Omega R\,, \quad \kappa =2\pi\nu/\bar{\mu}\,.\label{Nvvort}\ee
Here   $N_v^{\rm rig}=R^2/R_L^2$ is the  total number of vortices inside the vessel
of the internal radius $R$,
which should be formed at given $\Omega$ in order the interior of the vessel would rotate as a rigid body together with the walls, and $n_v=1/(\pi R_L^2)$ is the
corresponding
number of vortices per unit area. Thus
 distance between vortices
$R_L=\sqrt{\nu}/\sqrt{\bar{\mu}\Omega}\,,$
decreases with increasing $\Omega$.

The  energy gain due to the rigid-body rotation of  the  lattice of vortices mimicking the rotation of the vessel is given by  \cite{Tilly-Tilly},
\be
\delta {\cal{E}}\simeq N_v^{\rm rig} [{\cal{E}}_{\rm kin}^{(1)}(R_L) -L^v(R_L,\nu)\Omega]
\,.\label{chirvortengain}\ee
This result  is obtained within a simplifying assumption of a uniform distribution of vortices \cite{BaymChandler1983}.  A more accurate result   computed for the triangular lattice \cite{Tkachenko,FischerBaym2003}, differs only by a factor $\frac{\pi}{2\sqrt{3}}\simeq 0.91$  from  that found for the uniform approximation. Also, following simplifying consideration of Ref. \cite{Tilly-Tilly} we disregarded a small difference of  the rotation angular velocity of the vortex lattice, $\omega$, from that of the vessel $\Omega$. For $R{m}\gg 1$ this difference proves to be a tiny quantity, cf. \cite{Voskresensky:2023znr}.
Minimization of  (\ref{chirvortengain})  yields
\be N_v^{\rm rig} L_v(R_L,\nu)=N_v^{\rm rig}\pi n_{\pi} \nu^2 d_z/(2\bar{\mu}\Omega)
\,.\label{Lvortlat}\ee
 Setting (\ref{Lvortlat}) to (\ref{chirvortengain}) and using (\ref{Nvvort})
 we obtain the equilibrium  energy
 \be \delta {\cal{E}}
 \simeq n_{\pi}  \nu \Omega \pi R^2 d_z[\ln (R_L/r_0) -1/2]
 \,.\label{latdeltae}
 \ee
Minimum of $\delta {\cal{E}}$ corresponds to $\nu =1$.

Thus,  we demonstrated that at the rotation frequency $\Omega >\Omega_{c1}^\lambda$  in the rotating vessel filled by a pion gas (in our case at $T=0$)  there may appear charged pion vortices, which at subsequent increase of $\Omega$ (for $\Omega>2\Omega_{c1}^\lambda$) may form  the lattice mimicking the rigid-body rotation.

 With a further increase of the  rotation frequency
the lattice can be destroyed. The  minimal distance  $R_L \sim r_0$ at a dense packing of vortices  corresponds to the number of vortices per unit area $n_v\sim 1/(\pi r_0^2)$ in Eq. (\ref{Nvvort}) and  the maximum rotation frequency is
$\Omega \simeq \Omega_{c2}\sim 1/(r_0^2 {m})\,.$
For  $\Omega >\Omega_{c2}$, the $\phi$ vortex-state should disappear completely.
Note that for an extended rotating system the value $\Omega_{c2}R\gg 1$, however now it may not contradict to causality since the system may consist of independently rotating vortices. Note also that in cold atomic gases breakup of lattice occurs for $\Omega>\Omega_h\sim \Omega_{c2}r_0/R\ll \Omega_{c2}$ when in the center of vessel arises a hole,  cf. \cite{FischerBaym2003}.

\section{Some consequences}

{\em Rotating supercharged nuclei  and nuclearites.}\,
Let us consider a supercharged  nucleus or a piece of nuclear matter (nuclearite) of a large atomic number $A\simeq 2Z$  for
$ n_p=n_0/2$,   $Z$ is the proton charge of the nucleus and $n_0$ is the nuclear saturation density, $n_0\simeq m^3_\pi/2$. For  $V_0 \sim Ze^2/R> m_\pi$ the ground state $\pi^-$ energy level reaches zero and in reactions, e.g. $n\to p+\pi^-$,
there may appear the charged $\pi^-$ condensate.    For $Z|e^3|\gg 1$ the charge of protons is  screened by $\pi^-$ condensate at least to the value of a surface charge $Z_s\lsim Z/(Z|e^3|)^{1/3}$, cf. \cite{Migdal:1977rn}.  The surface energy term is small and
the  total energy  is given by
${\cal{E}}\simeq{\cal{E}}_{\pi,V}+{\cal{E}}_{A}\,,$ ${\cal{E}}_{A}\simeq -32Z$ MeV.
Let for simplicity  $\lambda \to 0$ and $m^*(n_0)$ is the effective mass of the pion in muclear matter at  $n=n_0$. Then in case of a resting nuclearite one obtains
${\cal{E}}\simeq (m^* -32 \,\rm{MeV})Z\,.$
Thus, if  $m^*(n_0)$ were $<32 \,\rm{MeV}$ or if there existed spinless charged bosons of such a mass, there would exist nuclearites and nuclei-stars  bounded by nuclear and electromagnetic forces, cf. \cite{Voskresensky1977,
MSTV90}. However, in spite of the attractive pion-nucleon interaction, pions have a larger effective mass at $n\sim n_0$, cf.
\cite{MSTV90,Voskresensky:1993ud}.

In case of a rotating supercharged nucleus, using (\ref{largenulim}), (\ref{bosenrot}), for $\epsilon_{1,\nu}=\mu=0$, and  for $\sqrt{m_\pi R}\gg c_1\gg 1$ we have $V_0\simeq m_\pi (1-\Omega R)c_1 $ and
\be {\cal{ E}}-{\cal{E}}_{in}\simeq [m_\pi (1 -\Omega R)c_1  -32 \mbox{MeV}]Z\label{freesuperchargedRot}.\ee
Vortex condensate state appears for ${\cal{ E}}-{\cal{E}}_{in}<0$.
This estimate is changed only a little in
 a realistic case,   $\lambda \sim 1$ within the $\lambda|\phi|^4$ model. Thus a charged pion  supervortex should be formed
 at least for $\Omega$ close to $1/R$. Additionally, we should notice that the rotating charged nucleus forms a magnetic field, which presence  still improves conditions for formation of the supervortex \cite{Voskresensky:2023znr}.

Thus in case of  a rotating  cold  piece of the nuclear matter
it might be profitable to form a charged pion   vortex field, which will  for a while stabilize the system. The kinetic energy of such a rotating nuclear system is then lost on a long time scale via a surface electromagnetic radiation. For very large number of baryons, $A$, such a radiation is strongly suppressed, e.g. pulsars radiate their energy during $\gsim 10^6$yr.

{\em Pion vortices in heavy-ion collisions.}\,
In heavy-ion collisions at LHC, RHIC conditions typical parameters of the pion fireball estimated in the resonance gas model \cite{Stachel} are: the temperature $T\simeq 155$ MeV, the volume is $5300$fm$^{3}$, the $\pi^{\pm}$ density is $n_{\pi}\simeq m^3_\pi$ and  we estimate the  electric potential as  $V_0\sim Ze^2/R\sim  0.2 m_\pi$ for central collisions, where $Z$ is now the  charge of the fireball. For peripheral collisions typical values of $V_0$ can be  larger.

Estimates   performed for peripheral heavy-ion collisions at $\sqrt{s}=200$ GeV give for the  rotation angular momentum values $L_z\lsim 10^6$ that yields $\nu\lsim 10^3$. Measured global polarization gave for vorticity $\Omega_{\rm exp}(200\rm{GeV}) \simeq 0.05 m_\pi$, cf. \cite{Xu-Guang Huang,Adamczyk}. Taking for size of the overlapping region of colliding nuclei $R = 10$ fm we get $\Omega <\Omega_{\rm caus}=1/R\simeq 0.14 m_\pi$.
 After the chemical freeze out up to the thermal freeze out, at $T_{\rm th}<T(t)\lsim m_\pi$,  the pion number can be considered dynamically fixed. Estimates show that  $T_{\rm{BEC}}>T_{\rm th}$, where $T_{\rm{BEC}}$ is the critical temperature of the pion Bose-Einstein condensation \cite{Voskresensky1994,Voskresensky1996,Kolomeitsev:2019bju}.   Thus  we may expect $\Omega\nu\gg  m_\pi$. A rough estimate yields that $\Omega_{c1}< \Omega_{\rm exp}(200\rm{GeV})$, whereas $\Omega_{c2}\sim m^2_\pi$. Thus the fireball   can be stabilized for a while by the formed pion supervortex or the lattice of vortices.  Also at these conditions one may think about occurrence of the  charged kaon  vortices.

 {{It is believed that  the spin polarization of  particles emitted in heavy-ion collisions is  induced by the coupling of the  angular momentum produced by colliding nuclei  and the spin of particles distributed in the matter.
  Nucleons participate  in production of strange particles,  e.g., $\Lambda$ hyperons.  The polarization of the $\Lambda$  is measured, cf. \cite{ITS}.
  Being formed,  pion vortices  absorb a part of the angular momentum.  At the freeze out they return part of the angular momentum back to baryons affecting
   $\Lambda$ polarization.
    Also, pion and baryon momentum
    distributions for particles  involved in  the vortex structures  should be different from the ordinary thermal  distributions. }}

{\em Superfluidity of baryon Cooper pairs and boson condensates in neutron stars.}\,
Nucleons in pulsars form  neutron-neutron and proton-proton Cooper pairs, playing a role of boson excitations,  and  $\bar{\mu}\simeq 2m_N$, where $m_N $ is the nucleon mass. Vortices in nucleon superfluids form the lattice, which mimics the rigid-body rotation of the matter. Besides the nucleons, the hyperons may appear in the interiors of neutron stars with the mass $M\gsim 1.4 M_{\odot}$, forming Cooper pairs.
Taking $\nu =1$ and using Eq. (\ref{Nvvort}) one gets estimation $n_v\simeq 6.3\cdot 10^3 (P/\rm{sec.})^{-1}$ vortices$/$cm$^2$ provided rotation period $P$ is measured in seconds, cf.  \cite{Sauls1989}. Then for the Vela pulsar, $P\simeq 0.083$ sec, the distance between vortices is $~4\cdot 10^{-3}$ cm.
The charged pion condensate superfluid also can be formed in neutron star interiors \cite{Voskresensky:1980nk}. Employing $\bar{\mu}\simeq m_\pi$  we estimate the density of pion vortices as  $n_v\simeq 5\cdot 10^2 (P/\rm{sec.})^{-1}$ vortices$/$cm$^2$. Similarly, the kaon \cite{Kolomeitsev:2002pg} and $\rho^{\pm}$ condensates \cite{Voskresensky:1997ub}  may form lattice structures.

{\em Rotating  vessel.}\,
Let  $n_p=0$. In case when an ideal  rotating vessel is placed inside a cylindrical co-axial charged capacitor (with plates placed at $r=R_{ex}$ and $r=R_{in}$ for $R_{ex}>R_{in}>R_{>}$)  when $Z_{in}$ is the charge placed on the internal surface of the capacitor and $-Z_{ex}$ is the charge of the external surface, the electric field  between plates is $E(r>R_{in})=-2Z_{ex} /(  r d_z)$ and $V(r=R_{in})=R_{in} E(R_{in})\ln (R_{ex}/R_{in})$. For $\rho_\pi d_z \pi R^2\ll Z_{ex}$ we have $V(r<R)\simeq V(r=R_{in})$.
For a surface charge density $\sigma\sim 10^4$v$/$cm and $R_{in}\sim 10$\,m, $R_{ex}\gsim (2-3)R_{in}$,  one gets $V_0>m_\pi$ and the supervortex arises even for $c_1\ll 1$.

In case of the pion gas, with the help of Eq.  (\ref{Nvvort}) we may estimate number of vortices in the lattice at the rigid-body rotation
   $N_v =\bar{\mu} \Omega R^2/\nu <\bar{\mu} (R/R_{>})^2 R_{>}/\nu\,,$
for $\Omega <1/R_{>}$, $1/R_{>}=3\cdot 10^6$Hz for  $R_{>}=10$m. Then, with $R=1$cm and $\bar{\mu} \sim m_\pi$  we estimate $N_v<10^{10}/\nu$, $R_L\gsim \sqrt{\nu/\Omega[\rm Hz]} 10^{-2}$cm.

{\em Formation of vortices in magnetic field.}\,
In  \cite{Zahed} rotation of the vacuum of non-interacting charged pions was considered in presence of a  strong  external uniform constant magnetic field $H$.   The number of permitted states  is given by $N=|e|HS/(2\pi)=|e|HR^2/2$ and for $|e|H< 2/R^2$ we have $N<1$.  Thus results \cite{Zahed} do not describe  the  case $H\to 0$, which we have studied above.

 Uniform magnetic field inside the rotating ideal vessel can be generated, if the vessel is put inside a solenoid or if we deal with  the charged rotating cylindrical capacitor. In the latter case  simple estimate shows that for  $R=1$ cm it is  sufficient to switch on a tiny  external field $|e|H>10^{-8}$G
in order to get $N>1$ and thereby to overcome the problem with absence of the  solution $\mu =0$ of
Eq. (\ref{grlev}) at $V_0=0$. Now, at $H\neq 0$, $N>1$, dispersion equation renders
$\epsilon_{1,\nu}=-\Omega\nu -V_0+\sqrt{{m}^2+|e|H}$
and instead of Eq. (\ref{critOmvac}) we obtain
\begin{eqnarray} &
\nu \Omega_c^H=\nu {{\Omega (\epsilon_{1,\nu}=0)}} =-V_0+\sqrt{{m}^2 +|e|H} \,,
\label{critOmvacH}\end{eqnarray}
and $\nu \Omega_c^H\simeq -V_0+{m}$  for $|e|H\ll m_\pi^2$, compare with Eq. (\ref{Omcrsmall}), the latter is valid only for $c_1\ll 1$. The  dependence on $R$  disappeared, cf.  \cite{Zahed}. The degeneracy factor $0<\nu\leq N$.
Fields $H\lsim (10^5-10^7)$G  can be  generated at the terrestrial laboratory conditions. For $|e|H\sim 10^6$ G at $R=1$ cm we  have $N\sim
10^{14}$ and $\nu\lsim N$  for $c_1\lsim 10$ and $\Omega_c^H\lsim 10^{9}$ Hz.

{\em Injection of the proton gas in rotating vessel.}\,
In absence of the capacitor, in case of the rotation of the charge neutral empty  vessel, in which  an amount of heavy positively charged particles is injected (e.g. protons), the positive  charge density $n_p$ can be compensated by the produced  negatively charged pion vortex field, i.e. $|n_\pi|=\bar{\mu}|\phi|^2 \simeq n_p$.
  Maximum value of  $|\phi|^2$   at $m^2_\pi\gg eH\sim Ze^2\Omega/R\gg 1/R^2$ corresponds to $\bar{\mu}\simeq {m}_\pi$,
 and thus
 the minimum of the pion supervortex energy in the rotation frame is given by
 ${\cal{E}}_{\pi}=({m}_\pi-\Omega\nu) Z$ and   it  becomes negative for $\Omega>{m}_\pi/\nu$, for $c_1>1$.

\section{Concluding remarks}\label{concluding-sect}
In this work within the $\lambda|\phi|^4$ model we studied possibilities of the formation of the charged $\pi^{\pm}$   vortex fields in various systems:   in the rotating cylindrical empty vessel; in the vessel filled by the charged pion gas at the temperature $T=0$  (Bose-Einstein condensate) with a (dynamically) fixed   particle number; and in case of rotating nuclear systems. In  case of the vessel filled by a pion gas at $T=0$, an analogy  was elaborated with  cold Bose gases and the condensed  $^4$He. Various applications of the  results were discussed.


\end{document}